\documentclass[a4paper]{jpconf}
\usepackage{graphicx}
\usepackage{color}
\begin{document}
\title{Pion transition form factor through Dyson-Schwinger equations}

\author{Kh\'epani Raya}

\address{Instituto de F\'isica y Matem\'aticas, Universidad Michoacana de San Nicol\'as de Hidalgo\\ Edificio C-3,
Ciudad Universitaria, C.P. 58040, Morelia, Michoac\'an, M\'exico.}

\ead{khepani@ifm.umich.mx}

\begin{abstract}
In the framework of Dyson-Schwinger equations (DSE), we compute
the $\gamma^*\gamma\to\pi^0$ transition form factor, $G(Q^2)$. For
the first time, in a continuum approach to quantun chromodynamics
(QCD), it was possible to compute $G(Q^2)$ on the whole domain of
space-like momenta. Our result agrees with CELLO, CLEO and Belle
collaborations and, with the well-known asymptotic QCD limit,
$2f_\pi$. Our analysis unifies this prediction with that of the
pion's valence-quark parton distribution amplitude (PDA) and
elastic electromagnetic form factor.
\end{abstract}
\section{Introduction}
The neutral pion transition form factor is measured through
electron-positron scattering. Although the available data of CELLO
\cite{Behrend:1990sr}, CLEO \cite{Gronberg:1997fj}, Babar
\cite{Aubert:2009mc} and Belle \cite{Uehara:2012ag} collaborations
agree in the domain of $Q^2<10$ GeV$^2$, the Babar and Belle data
(the only data available above that point) are notoriously
different. Moreover, how possibly the Babar data can
reconcile with the asymptotic QCD limit, calculated
by Brodsky and Lepage in \cite{Lepage:1980fj}, i.e., $2f_\pi$,
remains unclear and unsatisfactory.
\\
\\
We have previous DSE input from \cite{Maris:2002mz}, but because
of the numerical methods developed by that time, it
was not possible to arrive at momentum scales larger than $Q^2>4$
GeV$^2$. Complete understanding of $G(Q^2)$ demands simultaneously
achieving correct asymptotic behavior, but also the essentially non perturbative Abelian anomaly,
$2f_\pi G(Q^2=0) = 1$. Such features are achievable in the
framework of DSEs. At the same time, we are able to connect our
results with that of the pion's PDA, \cite{Chang:2013pq}, and
elastic electromagnetic form factor, \cite{Chang:2013nia}.
The reference to our detailed published article is
 \cite{Raya:2015gva}. Most of the ingredients which have gone into this study are quite general and can be applied to other mesons and
processes. In particular, the study of
$\gamma^*\gamma\to \eta_c$ and $\gamma^*\gamma\to\eta_b$
is under way.

\section{The tools}
For any pseudoscalar meson ($PS$), the $\gamma^*\gamma \to PS$ is
expressed through
$\mathcal{T}_{\mu\nu}(k_1,k_2)=T_{\mu\nu}(k_1,k_2)+T_{\mu\nu}(k_2,k_1)$,
where the matrix element is:
\begin{equation}
T_{\mu\nu}(k_1,k_2)=\frac{e^2}{4\pi^2}\epsilon_{\mu\nu\alpha\beta}k_{1\alpha}k_{2\beta}\;G(k_1^2,k_2^2,k_1\cdot
k_2)\;.
\end{equation}
The photons' momenta are $k_1$ and $k_2$ and the meson's total
momentum is $P=k_1+k_2$ ($P^2=-m_{PS}^2$). At leading order
(rainbow-ladder truncation) in the systematic and symmetry
preserving DSE truncation scheme, one can write:
\begin{equation}
\label{eq:TFFeq}
T_{\mu\nu}(k_1,k_2)=\int \frac{d^4k}{(2\pi)^4}i\mathcal{Q}\chi_\mu(l,l_1)\Gamma_{PS}(l_1,l_2)S(l_2)i\mathcal{Q}\Gamma_\nu(l_2,l)\;.
\end{equation}
Here $\mathcal{Q}$ is the quark charge operator ($e\;
\mbox{diag}[2/3,-1/3]$ for neutral pion), $l_1=l+k_1$,
$l_2=l-k_2$, where the kinematic constraints are: $k_1^2=Q^2$,
$k_2^2=0$, $k_1\cdot k_2 = - (m_{PS}^2+Q^2)/2$ and
$P^2=-m_{PS}^2$. The quark propagator, $S(k)$, and the
Bethe-Salpeter amplitude (BSA), $\Gamma_{PS}(q_1,q_2)$, are
obtained by solving the corresponding DSEs and the BSEs. On the
other hand, the quark-photon vertex is constructed via the
\emph{gauge technique}, \cite{Delbourgo:1977jc}. We
give the details in the following subsections.

\subsection{Quark propagator and Bethe-Salpeter amplitudes}
Most generally, the quark propagator is written as
$S(p) = -i \gamma \cdot p \sigma_v(p^2) + \sigma_s(p^2)$, while
the BSA is:
\begin{equation}
\Gamma_{PS}(k;P)=\gamma_5[i\;E_{PS}(k;P)+\gamma\cdot P F_{PS}(k;P)+ \gamma\cdot k\; k\cdot P G_{PS}(k;P)+\sigma_{\mu\nu}k_\mu P_\nu H_{PS}(k;P)]\;,
\end{equation}
The corresponding DSE and BSE, in the rainbow-ladder truncation,
are:
\begin{eqnarray}
\label{eq:DSErainbow}
S^{-1}(p) &=& \mathcal{Z}_{2F} (S^0)^{-1}(p) + \mathcal{Z}_{1F} \int \frac{d^4q}{(2\pi)^4}G(p-q) D_{\mu\nu}^0(p-q,\mu)\frac{\lambda^a}{2}\gamma_\mu S(q,\mu) \frac{\lambda^a}{2}\gamma_\nu,\\
\label{eq:DSEladder}
\Gamma_{PS}(p,P) &=& -\int \frac{d^4q}{(2\pi)^4} \frac{G((p-q)^2)}{(p-q)^2} \frac{\lambda^c}{2}\gamma_\mu S^a(q+\eta P) \Gamma_{PS}(q,P) S^b(q-(1-\eta P)) \frac{\lambda^c}{2}\gamma_\nu\;,
\end{eqnarray}
with $G(p-q)$ being the effective coupling described
in \cite{Qin:2011dd}. Once we obtain the solution of Eq.
(\ref{eq:DSErainbow}), we can parameterize it in the
form of a complex conjugate pole representation:
\begin{equation}
\label{eq:quarkparam}
S(p) = \sum_{j=1}^N \left( \frac{z_j}{i\gamma \cdot p+m_j}+\frac{z_j^*}{i\gamma \cdot p+m_j^*}\right)\;,
\end{equation}
constrained to the ultraviolet conditions of free quark propagator; $N=2$ is accurate enough for our purposes. We
then solve Eq. (\ref{eq:DSEladder}), and parameterize its
solutions with the perturbation theory integral
representation (PTIR):
\begin{eqnarray}
\nonumber
\mathcal{F}(k;P)&=&\mathcal{F}^i(k;P) + \mathcal{F}^u(k;P)\;,\\
\nonumber
\mathcal{F}^i(k;P)&=& c_{\mathcal{F}}^i \int_{-1}^1 dz \;\rho_{\nu^i_\mathcal{F}}(z) [ a_\mathcal{F} \hat{\Delta}^4_{\Lambda_\mathcal{F}^i}(k_z^2) + \bar{a}_\mathcal{F} \hat{\Delta}^5_{\Lambda_\mathcal{F}^i}(k_z^2) ]\;,\\
\nonumber
E^u(k;P)&=& c_{E}^u \int_{-1}^1 dz \; \rho_{\nu^u_E}(z) \hat{\Delta}^{1+\alpha}_{\Lambda_E^u}(k_z^2) \;,\\
\nonumber
F^u(k;P)&=& c_{F}^u \int_{-1}^1 dz \; \rho_{\nu^u_F}(z) k^2 \Lambda_F^u \Delta^{2+\alpha}_{\Lambda_F^u}(k_z^2) \;, \\
\nonumber
G^u(k;P)&=& c_{G}^u \int_{-1}^1 dz \; \rho_{\nu^u_G}(z) \Lambda_G^u \Delta^{2+\alpha}_{\Lambda_G^u}(k_z^2) \;,
\label{eq:BSAparam}
\end{eqnarray}
where $\mathcal{F}(k;P) = E,F,G$ and $i,u$ stand for IR and UV; $\hat{\Delta}_\Lambda(s) = \Lambda^2 \Delta_\Lambda(s)$, $\Delta_\Lambda(s) = (\Lambda^2+ s)^{-1}$, $k_z^2 = k^2 + z k \cdot P$, $\bar{a}_E = 1- a_E$, $\bar{a}_F = 1/\Lambda_F^i-a_F$, $\bar{a}_G = [1/\Lambda_G^i]^3-a_G$. The parameters $a$, $c$, $\Lambda$, $\alpha$ and  $\nu$ are fitted to the numerical data; $\alpha$ simulates the logarithmic UV behavior, characteristic of QCD, and $\nu$ defines the spectral density
\begin{equation}
\rho_\nu(z) = \frac{1}{\sqrt{\pi}}\frac{\Gamma(3/2+\nu)}{\Gamma(1/2+\nu)}(1-z^2)^\nu\;.
\end{equation}
The amplitude $H(k;P)$ has only a very small impact
on the final results and can safely be neglected. Such
representations have a quadratic form in the denominator, which
will be practically useful in the computation of Eq.
(\ref{eq:TFFeq}).

\subsection{Quark-photon vertex}
PTIRs are not available yet for the quark-photon vertex. Instead,
we use the following {\em ansatz} for the unamputated vertex:
\begin{eqnarray}
\chi_\mu(k_f,k_i)&=& \gamma_\mu \Delta_{k^2 \sigma_V} + [s \gamma\cdot k_f \gamma_\mu \gamma \cdot k_i
+ \bar{s}\gamma\cdot k_i \gamma_\mu \gamma \cdot k_f]\Delta_{\sigma_V}\nonumber\\
&+&[s(\gamma\cdot k_f \gamma_\mu + \gamma_\mu \gamma \cdot
k_i)+\bar{s}(\gamma\cdot k_i \gamma_\mu + \gamma_\mu \gamma \cdot
k_f)]i\Delta_{\sigma_S}\;, \label{eq:vertexAnsatz}
\end{eqnarray}
where $\Delta_F=[F(k_f^2)-F(k_i^2)]/[k_f^2-k_i^2]$, $q=k_f-k_i$
and $\bar{s}=1-s$. This vertex {\em ansatz} is obtained using the
\emph{gauge technique}. By construction, it
satisfies the longitudinal Ward-Green-Takahashi identity, is free
of kinematic singularities, reduces to the bare vertex in the
free-field limit, and has the same Poincar\'e transformations
properties as the bare vertex.
\\
\\
Up to transverse pieces associated with the scalar $s$,
$\chi_{\mu}(k_f,k_i)$ is equivalent to $S(k_f)
\Gamma_{\mu}S(k_i)$. Nothing material would be gained herein by
making them identical because any difference is power-law
suppressed in the ultraviolet; but computational effort would
increase substantially. We define $s$ as follows:
\begin{equation}
\label{eq:sDef} s = 1+s_0 \; {\rm Exp}[-\mathcal{E}_{PS}/M_E]
\end{equation}
where $\mathcal{E}_{PS}=\sqrt{Q^2/4+m_{PS}^2}-m_{PS}$ is the
\emph{Breit-frame} energy of the pseudoscalar and, $M_E = \{ p |
p^2 = M^2(p^2),\;p^2>0 \}$ is the Euclidian constituent-quark
mass. The parameter $s\neq1$ has to do with the value of
$G(Q^2)$ in a neighborhood of $Q^2=0$. In the case of the pion,
owing to the Abelian anomaly, it is impossible to simultaneously
conserve the vector and axial vector currents associated with Eq.
(\ref{eq:TFFeq}), but, with a proper choice of $s_0$ in Eq.
(\ref{eq:sDef}), vector currents are conserved and
the Abelian anomaly is satisfied. With everything expressed in
terms the same functions, we will now see how the appropriate parameterizations of $S(p)$ and $\Gamma_{PS}(k,P)$
allow us to compute $G(Q^2)$ in the whole range of space-like
momenta.

\section{The Calculation}
 All elements of Eq. (\ref{eq:TFFeq}) are written in terms of $S(p)$ and $\Gamma_{PS}(k,P)$, expressed as complex
 conjugate pole representation or PTIRs, respectively. Computation of $G(Q^2)$ reduces to the task of summing a series of terms,
 all of which involve a single four-momentum integral.
\\
\\
As we saw before, our quark-photon vertex construction allow us to
satisfy one of the constraints of the transition form factor,
namely, the conservation of vector current (ensuring
the Abelian anomaly is correctly recovered for the pion). In other words, this constraint fixes the value of
$G(Q^2=0)$. One should also understand the connection of $G(Q^2)$
with the meson's PDA and its evolution with the factorization
scale of QCD. As the same PDA also governs the $Q^2$
dependence of the pion electromagnetic form factor, we will have a
unified prediction of both the form factors in the asymptotic
limit of QCD. We detailed the algorithm below.

\subsection{The algorithm}
Because of the representations employed for $S(p)$ and
$\Gamma_{PS}(k,P)$, the denominator in every term is a product of
$l$-quadratic forms. We perform Feynman parameterization. After a
proper change of variables, all the momentum integrals can be
solved analytically. Once we solved the four-momentum integrals,
one computes a finite number of simple integrals, namely, the ones
over the Feynman parameters and $z$, the spectral integral. The
complete result for $G(Q^2)$ is obtained after summing the series.
\\
\\
The peculiar perturbation theory like
parameterizations and the procedure employed are the reason we
are able to compute $G(Q^2)$, for arbitrarily large $Q^2$
space-like momenta, for the first time in a continuum approach
directly connected to QCD. For a complete explanation and more
detail, please consult our published article
\cite{Raya:2015gva}. This procedure has also previously been
employed to compute the pion's elastic electromagnetic form factor
and pion's PDA \cite{Chang:2013pq,Chang:2013nia}.

\subsection{Pion's PDA and its evolution}
Various studies indicate that the pion PDA is a broad
concave function, and at the resolution scales achieved so far, it
is far from being similar to the asymptotic PDA,
$\phi_{asy}(x)=6x(1-x)$. According to \cite{Chang:2013pq}, it is
defined by the expression:
\begin{equation}
f_\pi \phi(x;\zeta) = N_c \mbox{tr} Z_2(\zeta,\Lambda) \int
\frac{d^4k}{(2\pi)^4} \delta_n^u(k_\eta)\gamma_5 \gamma\cdot n
S(k_\eta) \Gamma_\pi(k_{\eta\bar{\eta}};P)S(k_{\bar{\eta}})\;,
\label{eq:PDA}
\end{equation}
where $N_c=3$; the trace is over spinor indices;
$Z_2(\zeta,\Lambda)$ is the quark wave-function renormalisation
constant; $\delta_n^u(k_\eta):=\delta(n\cdot k_\eta - u n\cdot
P)$, with $n^2=0$, $n\cdot P = -m_\pi$; and
$k_{\eta\bar{\eta}}=[k_\eta+k_{\bar{\eta}}]/2$, $k_\eta = k + \eta
P$, $k_{\bar{\eta}}=k-(1-\eta)P$, $\eta \in [0,1]$. The way it
evolves towards its asymptotic limit is described through the
Efremov-Radyushkin-Brodsky-Lepage (ERBL) evolution equations
\cite{Lepage:1980fj,Efremov:1979qk}.
\begin{figure}[h!]
\begin{center}
\includegraphics[width=11cm]{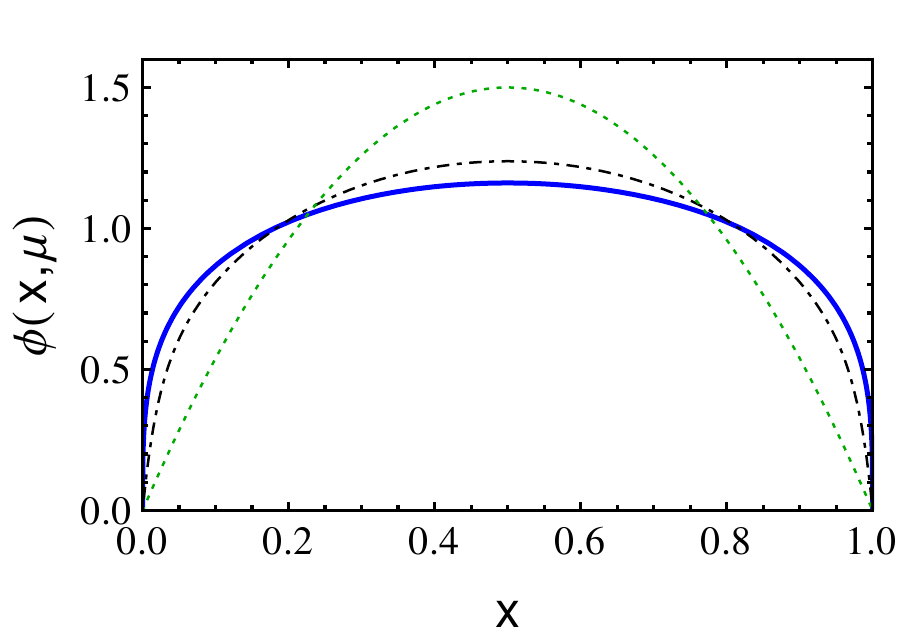}
\end{center}
\caption{\label{fig:PDA}\textbf{Pion's PDA}: [Solid, blue]
$\phi(x,\mu=2\;\mbox{GeV})$. [Dot-dashed, black] ERBL evolution to
$\phi(x,\mu=\sqrt{50}\;\mbox{GeV})$. [Green, dotted] Asymptotic
PDA, $\phi_{asy}(x)=6x(1-x)$. As we can see, the broad concave
PDA, $\phi(x,\mu=2\;\mbox{GeV})$, slowly approaches to its
asymptotic form with increasing momentum scale.}
\end{figure}
\\
\\
From figure \ref{fig:PDA}, we see that pion's PDA at resolution
scale $\mu=2$ GeV, $\phi(x,\mu = 2\; \mbox{GeV})$, slowly evolves to
its asymptotic form. Evolution enables the dressed-quark and
antiquark degrees of freedom, in terms of which the wave function
is expressed at a given scale, to split into less well-dressed
partons via the addition of gluons and sea quarks in the manner
prescribed by QCD dynamics. The connection of $\phi(x,\mu)$ with
$G(Q^2)$ is given by the leading twist expression for the
transition form factor:
\begin{equation}
G(Q^2)=4 \pi^2 f_\pi \int_0^1 dx
T_H(x,Q^2,\alpha(\mu);\mu)\phi(x,\mu)\;,
\end{equation}
where $T_H(\mu)$ is the photon-quark-antiquark scattering
amplitude at some scale $\mu$. One expects to reach the asymptotic
limit from below, otherwise one would have to explain why $G(Q^2)$
grows bigger and then decreases towards $2f_\pi$. QCD
is not known to have an additional scale to set in at a higher
$Q^2$ to make this happen. Only logarithmic corrections have a
minor role to play. We shall see that the evolution of
$\phi(x,\mu)$ is crucial in understanding the asymptotic behavior
of the pion transition form factor. Further details of pion's PDA can be found in Javier Cobos' contribution to this proceedings volume and in \cite{Chang:2013pq}.

\section{Results and Conclusions}
\subsection{Results}
\begin{figure}[h!]
\begin{center}
\includegraphics[width=11cm]{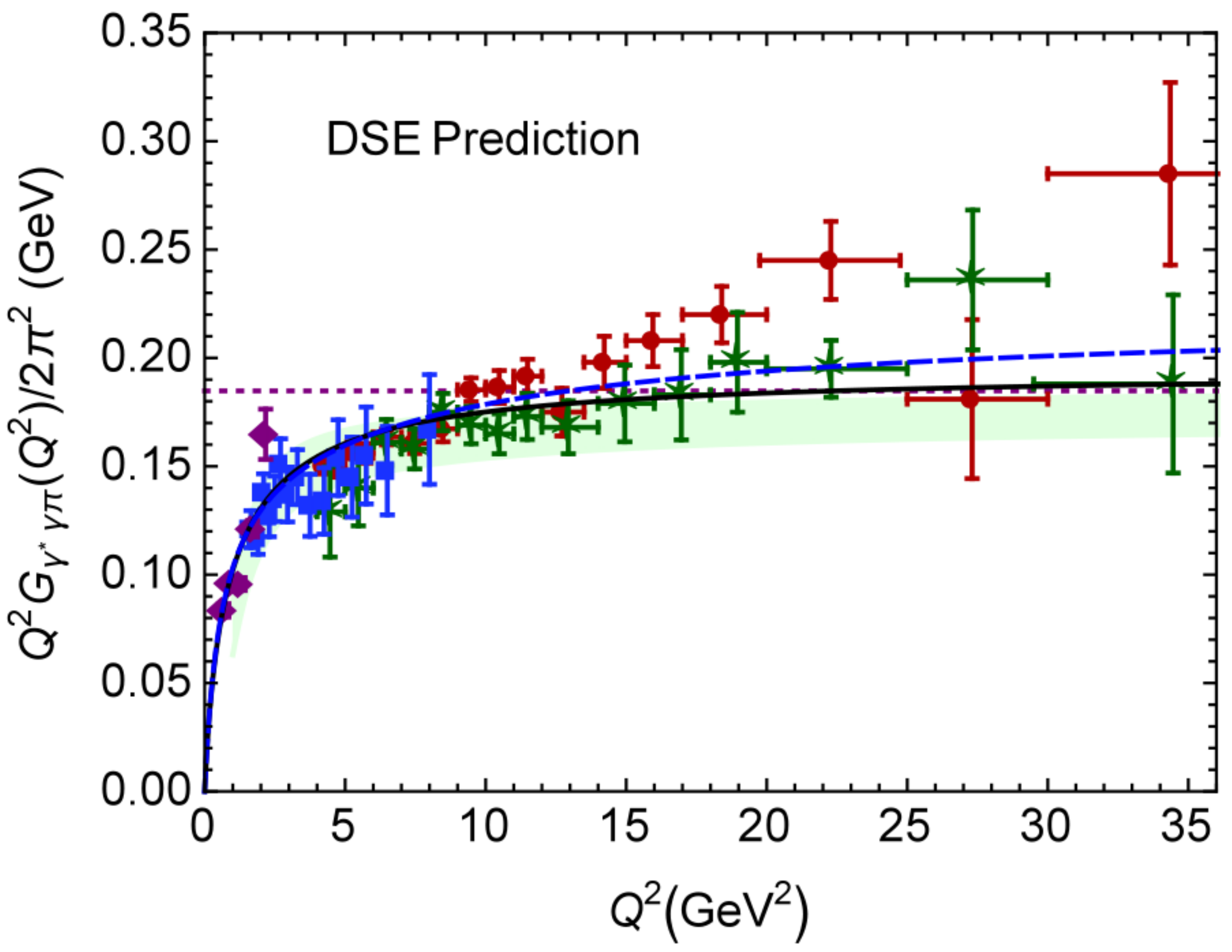}
\end{center}
\caption{\label{fig:TFF}\textbf{Pion's transition form factor}:
[Solid, black] DSE Prediction obtained with the ERBL evolution of
pion BSA. [Dashed, blue] DSE Prediction without evolution at
frozen scale $\mu=2\;\mbox{GeV}$. \textbf{Data:} [Circles, red] Babar
\cite{Aubert:2009mc}, [Diamonds, purple] CELLO
\cite{Behrend:1990sr}, [Squares, blue] CLEO
\cite{Gronberg:1997fj}, [Stars, green] Belle \cite{Uehara:2012ag}.
The (green) shaded band is described in \cite{Bakulev:2012nh}.}
\end{figure}

Our main result is depicted in figure \ref{fig:TFF}. We obtain an
interaction radius of $r_{\pi^0}=0.68$ fm, which is practically
identical to the one computed in \cite{Chang:2013nia} within the
 same scheme. The Abelian anomaly, namely $2f_\pi G(Q^2=0)
= 1$, is satisfied; the asymptotic limit, $2f_\pi$, is reached
from below except for a logarithmic miss-match. Our
result agrees fairly well with all available data below $Q^2<10$
GeV$^2$, and with Belle data at large $Q^2$ scales. However, it
fails to reconcile with the data reported by the Babar
collaboration.
\\
\\
From figure 2, we see that $G(Q^2)$ exceeds (logarithmically) the
asymptotic limit $2f_\pi$, although we expected such
limit to be reached from below.  However, as mentioned earlier,
the growth is only logarithmic, and at some point, it settles onto
the value $2f_\pi$. This discrepancy originates from
the failure of the rainbow-ladder (RL) truncation to
reproduce the complete set of gluon and quark splitting effects
contained in QCD and hence its inability to fully
express interferences between the anomalous dimensions of those
$n$-point Schwinger functions which are relevant in the
computation of a given scattering amplitude.

\subsection{Conclusions}
We describe a computation of the pion transition form factor, in
which all elements employed are determined by the solutions of the
QCD's DSEs, obtained in the RL truncation.
\begin{itemize}

\item The novel analysis techniques made it possible to compute
$G(Q^2)$, on the entire domain of space-like momenta, in a continuum framework directly connected to QCD.

\item Our work unifies the description and explanation of this
transition with the charged pion electromagnetic form factor and
its PDA.

\item

This enables us to demonstrate that a fully self-contained and
consistent treatment can readily connect a pion PDA, that is a
broad and concave function at the hadronic scale, with the
perturbative QCD prediction for the transition form factor in the
hard photon limit.

\end{itemize}
Full discussion and details are found in our work in
\cite{Raya:2015gva}.

\section{Acknowledgements}
I want to acknowledge the organizing committee for the financial support and my collaborators in \cite{Raya:2015gva} for fruitful discussions.

\end{document}